\documentclass[12pt]{iopart}

\expandafter\let\csname equation*\endcsname\relax
\expandafter\let\csname endequation*\endcsname\relax
\usepackage{amsmath}

\usepackage{iopams}   % loads amsfonts, amssymb and amsbsy
\usepackage{graphicx}
\usepackage{xcolor}

\newif\ifmarkrev 
\markrevtrue
\newcommand{\rev}[1]{\ifmarkrev{#1}\else#1\fi}
\newenvironment{revised}{\ifmarkrev\fi}{}

\newcommand{\vf}{v_{\mathrm{F}}}

\begin{document}

\title[Curvature as effective backgrounds for Dirac fermions]{From geometric
curvature to effective backgrounds: a dictionary for Dirac fermions in one,
two and three spatial dimensions}

\author{David Valenzuela$^1$ and Alfredo Raya$^{2,3}$}

\address{$^1$ Physics Solution, Casa 94, Pasaje 15, Calle La Ca\~nada,
Jardines de La Hacienda, Antiguo Cuscatl\'an, El Salvador}
\address{$^2$ Facultad de Ingenier\'{\i}a El\'ectrica, Universidad Michoacana
de San Nicol\'as de Hidalgo, Edificio Omega, Ciudad Universitaria,
Francisco J.\ M\'ujica s/n, Col.\ Fel\'{\i}citas del R\'{\i}o,
CP 58040, Morelia, Michoac\'an, Mexico}
\address{$^3$ Centro de Ciencias Exactas, Universidad del B\'{\i}o-B\'{\i}o,
Casilla 447, Chill\'an, Chile}

\ead{david.valenzuela@physicssol.com, alfredo.raya@umich.mx}

\begin{abstract}
\begin{revised}
Dirac equation on curved backgrounds can be rewritten as a flat-space equation in which geometry appears as a tensorial Fermi velocity together with a set of background fields. We derive this correspondence from a single Clifford-algebra identity and follow it through one, two, and three spatial dimensions. Once the spin connection is written in terms of Ricci rotation coefficients, only two contractions can occur. The trace produces a term with the index structure of a vector potential in every dimensionality; it enters with an imaginary factor  i and is precisely the term that keeps the operator Hermitian when the Fermi velocity depends on position. The totally antisymmetric contraction produces an object that depends on the number of indices available to the Levi-Civita symbol: a trace on a line, a pseudoscalar mass on a surface, and an axial vector potential in a volume. Neither contraction is invariant under local rotations of the frame, and we give the transformation laws. On a surface, the curl of the trace term can always be removed by a frame rotation, while its divergence is proportional to the Gaussian curvature. We apply the correspondence to a Gaussian bump, where the local density of states at the center falls below its flat value and obeys a closed form at low energy; to a wire, where a nonuniform lapse binds states inside the gap and a nonuniform velocity cannot; to stationary frames on a surface, where a circulating drag shifts the Dirac mass locally with either sign; and to twisted and rotating frames in a volume, where the axial term takes the form of a chiral chemical potential produced by the frame alone, or separates Weyl nodes in a rotating sample. We separate the geometric contribution from the lattice pseudogauge field that dominates the response of strained samples.\end{revised}
\end{abstract}


\submitto{\rev{Physica Scripta}}

\maketitle

% ---------------------------------------------------------------------------
\section{Introduction}
\label{sec:intro}
The low energy electronic structure of graphene is governed by a massless
Dirac operator, a fact that follows from the honeycomb lattice and was
already implicit in the band calculation of Wallace \cite{Wallace1947} long
before the material was isolated \cite{Novoselov2005,CastroNeto2009}. Real
samples, however, are never flat. Ripples appear spontaneously at finite
temperature, substrates impose corrugations, and deliberate straining bends
the sheet in controlled ways. The carriers then propagate on a curved
two dimensional (2D) manifold, and the question of how that geometry enters their
dynamics stops being academic.
Two traditions have grown around the question. One starts from the tight
binding description and treats the modulation of the hopping amplitudes as
the primary effect, which produces a valley dependent gauge potential whose
curl behaves as a magnetic field of large magnitude
\cite{Vozmediano2010,Guinea2010,Amorim2016}. Scanning tunneling
spectroscopy of strained nanobubbles has confirmed the prediction by
resolving pseudo Landau levels \cite{Levy2010}, and later work mapped the
fields directly \cite{Meng2013,Jiang2017}. The other tradition writes the
Dirac equation covariantly on the deformed surface
\cite{Kavalov1986,CortijoEPL2007,Cortijo2007,Mesaros2010} and asks what the
vielbein and the spin connection do to the spectrum, an approach that has
been pushed far enough to propose graphene as a setting for quantum field
theory in curved spacetime \cite{Iorio2011,IorioLambiase2012,Gallerati2021}.
The apparent tension between \rev{the two} descriptions was resolved by de Juan,
Sturla and Vozmediano, who showed that both lead to a space dependent Fermi
velocity \rev{of the same form} \cite{deJuan2012}, a
conclusion later supported by spectroscopy on corrugated monolayers~\cite{Yan2013}.

What makes the covariant route attractive is not elegance but arithmetic. A
curved manifold forces one to carry Christoffel symbols, a spin connection
and a position dependent Clifford algebra through every calculation. If the
same physics can be written as a flat space Dirac equation with a tensorial
Fermi velocity and a few background fields, then the whole apparatus of flat
space quantum field theory becomes available: ordinary perturbation theory,
scattering techniques, functional determinants, factorization methods for
bound states. Problems that look forbidding in curved coordinates, such as
the spectrum of a graphene wormhole \cite{Gonzalez2010}, become problems of a
kind that has been solved many times.
The translation itself, though, is usually presented for a particular
geometry and in a particular dimension. The leftover piece of the spin
connection is then read off case by case, and its nature tends to be
asserted rather than derived. That is a gap worth closing, because
\rev{three dimensional (3D)} Dirac and Weyl materials \rev{\cite{Armitage2018}}
raise the same questions and there is no reason to expect the 2D answer to
carry over unchanged.

\begin{revised}
Parts of the construction are not new, and it is fair to say so at the
outset. The splitting of the spin connection into a trace and a totally
antisymmetric piece is standard in the study of Dirac fields in Riemann-Cartan
spaces, where only the axial part of the torsion couples to the spinor
\cite{HehlDatta1971,Shapiro2002}. In graphene, the trace term is the
correction that keeps the Hamiltonian Hermitian when the Fermi velocity
depends on position \cite{deJuan2012,deJuan2013}. What we have not found in
the literature is a treatment that follows the antisymmetric contraction
across dimensions, states which background it becomes in each case, and keeps
track of how every entry of the correspondence changes with the choice of
frame. Those are the aims of this work, together with a set of worked
applications in which the correspondence produces definite results.
\end{revised}

\begin{figure}[htb]
\centering
\includegraphics[width=0.95\textwidth]{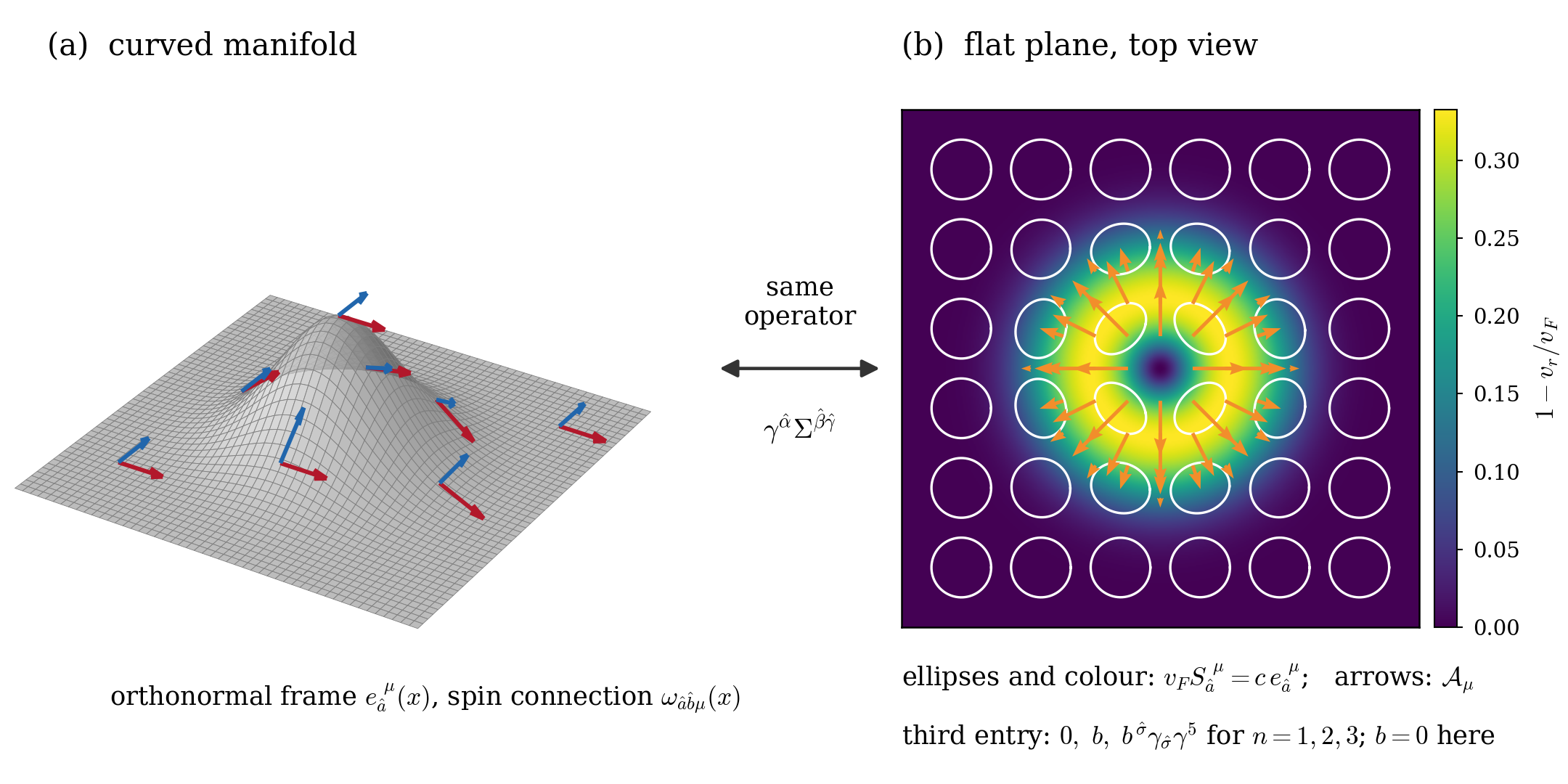}
\caption{Schematics of the correspondence between the Dirac equation on a
curved manifold and a flat plane with background fields.}
\label{fig:correspondence}
\end{figure}

The present work derives the correspondence from a single identity of the
Clifford algebra and follows it through \rev{one, two and three} spatial
dimensions, as illustrated in Figure~\ref{fig:correspondence}. The outcome
is that the trace of the Ricci rotation coefficients always produces \rev{a
term with the structure of} a vector potential, while the totally
antisymmetric contraction produces an object whose character is dictated by
how many indices the Levi-Civita symbol has room for: nothing on a line,
\rev{a pseudoscalar mass} on a surface, an axial vector potential in a volume.
\rev{We have organized the rest of the manuscript as follows.} Section
\ref{sec:setup} fixes the geometric setting, section \ref{sec:dictionary}
establishes the correspondence \rev{and its behavior under changes of frame},
section \ref{sec:applications} \rev{works out membranes, a Gaussian bump, a
wire, stationary frames on a surface, and twisted and rotating frames in a
volume}, and section
\ref{sec:scope} delimits what the correspondence does and does not describe.

% ---------------------------------------------------------------------------
\section{Dirac fermions on a curved background}
\label{sec:setup}

Two families of coordinates are needed throughout. The set
$\{x^{\mu}\}_{\mu=0}^{n}$ labels points on the curved manifold, while hatted
indices refer to a local orthonormal frame in which the metric takes the
Minkowski form $\eta^{\hat\alpha\hat\beta}=\mathrm{diag}(1,-1,\dots,-1)$ in
$n+1$ dimensions. \rev{Since the manuscript is concerned with Dirac
materials, the limiting speed of the analogue metric is the Fermi velocity
from the start, and the time coordinate is $x^{0}=\vf t$.} The vielbein
$e^{\hat\alpha}_{\ \mu}$ carries one index of each kind and relates the two
descriptions through
\begin{equation}
g_{\mu\nu}=\eta_{\hat\alpha\hat\beta}\,
e^{\hat\alpha}_{\ \mu}e^{\hat\beta}_{\ \nu},
\qquad
e^{\hat\alpha}_{\ \mu}e_{\hat\beta}^{\ \mu}=\delta^{\hat\alpha}_{\hat\beta}.
\label{eq:vielbein}
\end{equation}
\rev{The correspondence derived below holds for any orthonormal frame.
The individual backgrounds it produces, however, are not frame invariant, and
section~\ref{sec:frames} gives their transformation laws. A convenient
choice of frame shortens the algebra considerably in the applications of
section~\ref{sec:applications}.}

A spinor propagating on such a background obeys
\begin{equation}
\left(e_{\hat\alpha}^{\ \mu}\gamma^{\hat\alpha}\rev{\vf}P_{\mu}+m\rev{\vf^{2}}\right)\Psi=0,
\label{eq:dirac}
\end{equation}
with the generalized momentum $P_{\mu}=-i\hbar(\partial_{\mu}+\Omega_{\mu})$.
The piece that distinguishes equation (\ref{eq:dirac}) from its flat space
counterpart is the affine spinor connection
\begin{equation}
\Omega_{\mu}=\frac{1}{2}\,\omega_{\hat\alpha\hat\beta\mu}
\Sigma^{\hat\alpha\hat\beta},
\qquad
\Sigma^{\hat\alpha\hat\beta}
=\frac{1}{4}\left[\gamma^{\hat\alpha},\gamma^{\hat\beta}\right],
\label{eq:omega}
\end{equation}
built from the spin connection
\begin{equation}
\omega^{\hat\alpha}_{\ \hat\beta\mu}
=-e_{\hat\beta}^{\ \nu}\left(\partial_{\mu}e^{\hat\alpha}_{\ \nu}
-\Gamma^{\lambda}_{\mu\nu}e^{\hat\alpha}_{\ \lambda}\right),
\label{eq:spinconn}
\end{equation}
where $\Gamma^{\lambda}_{\mu\nu}$ are the Christoffel symbols of the
Levi-Civita connection. Metric compatibility forces
$\omega_{\hat\alpha\hat\beta\mu}=-\omega_{\hat\beta\hat\alpha\mu}$, and this
antisymmetry in the frame indices is used repeatedly below.

The spin connection is an awkward object precisely because it mixes index
types. \rev{Two} of its labels live in the frame and one lives on the
manifold, so it transforms partly as a tensor and partly as a connection. The
remedy is to push the last index into the frame as well,
\begin{equation}
\omega_{\hat\alpha\hat\beta\hat\gamma}
\equiv e_{\hat\gamma}^{\ \mu}\,\omega_{\hat\alpha\hat\beta\mu},
\label{eq:ricci}
\end{equation}
which defines the Ricci rotation coefficients
\cite{Nakahara2003,BirrellDavies1984}. Because the vielbein is invertible by
construction, equation (\ref{eq:ricci}) loses no information and can be read
backwards, $\omega_{\hat\alpha\hat\beta\mu}
=e^{\hat\gamma}_{\ \mu}\omega_{\hat\alpha\hat\beta\hat\gamma}$. What the
change of basis buys is that the resulting coefficients are ordinary
functions carrying \rev{three} frame indices, antisymmetric in the first pair,
and they are the only geometric data the Dirac operator responds to. Every
statement in section \ref{sec:dictionary} is a statement about how those
functions get distributed among the effective backgrounds.

% ---------------------------------------------------------------------------
\section{The dictionary}
\label{sec:dictionary}

The claim to be established is that equation (\ref{eq:dirac}) can be traded
for a flat space problem of the form
\begin{equation}
\left(\vf S_{\hat\alpha}^{\ \mu}\gamma^{\hat\alpha}
\left(p_{\mu}+ie\mathcal{A}_{\mu}\right)+m\vf^{2}+\mathcal{V}\right)\Psi=0,
\label{eq:model}
\end{equation}
written with the ordinary momentum $p_{\mu}=-i\hbar\partial_{\mu}$. Here
$\vf$ is the Fermi velocity, $S_{\hat\alpha}^{\ \mu}$ plays the role of a
strain matrix that makes that velocity tensorial and position dependent,
$\mathcal{A}_{\mu}$ \rev{is a term with the index structure of a vector
potential} and $\mathcal{V}$ collects whatever is left over. \rev{With the
analogue metric written in terms of $\vf$, comparison of equations
(\ref{eq:dirac}) and (\ref{eq:model}) identifies the strain matrix directly,}
\begin{equation}
\rev{S_{\hat\alpha}^{\ \mu}\equiv e_{\hat\alpha}^{\ \mu}.}
\label{eq:strain}
\end{equation}
An anisotropic, position dependent Fermi velocity is therefore not an extra
ingredient of the model but the inverse vielbein itself, which is consistent
with what the tight binding treatment of nonuniform strain produces
\cite{deJuan2012,OlivaLeyva2015} and with the direct spectroscopic evidence
for a spatially varying velocity in corrugated samples \cite{Yan2013}.

All the remaining work sits in the connection term. Contracting equation
(\ref{eq:omega}) with the Dirac structure and using (\ref{eq:ricci}) to
project the curved index gives
\begin{equation}
Q\equiv e_{\hat\alpha}^{\ \mu}\gamma^{\hat\alpha}\Omega_{\mu}
=\frac{1}{2}\,\omega_{\hat\beta\hat\gamma\hat\alpha}\,
\gamma^{\hat\alpha}\Sigma^{\hat\beta\hat\gamma}.
\label{eq:Q}
\end{equation}
Everything now hinges on the product $\gamma^{\hat\alpha}
\Sigma^{\hat\beta\hat\gamma}$, and that product is fixed by the Clifford
algebra alone. It splits into a piece proportional to a single gamma matrix,
with \rev{two} frame indices contracted by the metric, and a piece
proportional to the dual of the index triplet. In $3+1$ dimensions the
statement reads
\begin{equation}
\gamma^{\hat\alpha}\Sigma^{\hat\beta\hat\gamma}
=\frac{1}{2}\left(\eta^{\hat\alpha\hat\beta}\gamma^{\hat\gamma}
-\eta^{\hat\alpha\hat\gamma}\gamma^{\hat\beta}\right)
-\frac{i}{2}\,\varepsilon^{\hat\sigma\hat\alpha\hat\beta\hat\gamma}
\gamma_{\hat\sigma}\gamma^{5},
\label{eq:master}
\end{equation}
and appendix \ref{app:gammas} shows how it follows from the standard product
of \rev{three} gamma matrices. The first term is what survives when
\rev{two} of the \rev{three} indices are contracted, the second is what
survives when all \rev{three} are distinct. That division carries the whole
content of the dictionary, because the spin connection is antisymmetric in
its first \rev{two} labels and therefore admits exactly those \rev{two}
independent contractions with a \rev{three} index object.

It is convenient to name them once. The trace vector and its dual partner
are
\begin{equation}
a_{\hat\gamma}\equiv\omega^{\hat\alpha}_{\ \hat\gamma\hat\alpha}
=\eta^{\hat\alpha\hat\beta}\omega_{\hat\beta\hat\gamma\hat\alpha},
\qquad
b^{\hat\sigma}\equiv\varepsilon^{\hat\sigma\hat\alpha\hat\beta\hat\gamma}
\omega_{\hat\beta\hat\gamma\hat\alpha}.
\label{eq:ab}
\end{equation}
Neither object is exotic. The first is the contraction on which a Weyl
rescaling of the spinor acts, and the second is the totally antisymmetric part
of the anholonomy of the frame. With these definitions, inserting (\ref{eq:master})
into (\ref{eq:Q}) and using the antisymmetry of $\omega$ to fold the
\rev{two} metric terms into one leaves
\begin{equation}
Q=\frac{1}{2}\,a_{\hat\gamma}\gamma^{\hat\gamma}
-\frac{i}{4}\,b^{\hat\sigma}\gamma_{\hat\sigma}\gamma^{5},
\label{eq:Qsplit}
\end{equation}
so that the curved space equation (\ref{eq:dirac}) becomes
\begin{equation}
\left(\rev{\vf}\,e_{\hat\alpha}^{\ \mu}\gamma^{\hat\alpha}p_{\mu}
-\frac{i\hbar \rev{\vf}}{2}a_{\hat\gamma}\gamma^{\hat\gamma}
-\frac{\hbar \rev{\vf}}{4}b^{\hat\sigma}\gamma_{\hat\sigma}\gamma^{5}
+m\rev{\vf^{2}}\right)\Psi=0.
\label{eq:dirac4d}
\end{equation}
Comparison with (\ref{eq:model}) is now a matter of reading off coefficients.
The vector \rev{term} is the trace of the Ricci rotation coefficients pulled
back to the manifold,
\begin{equation}
\mathcal{A}_{\mu}=-\frac{\hbar}{2e}\,e^{\hat\gamma}_{\ \mu}a_{\hat\gamma},
\label{eq:Amu}
\end{equation}
with $e$ the electron charge.

\begin{revised}
Two properties of this term must be kept in view, because they decide how it
should be interpreted. It enters equation (\ref{eq:model}) with a factor of
$i$, so it is anti-Hermitian with respect to the flat measure and is not a
$U(1)$ gauge field: it cannot be removed by a change of phase, only by a
change of normalization, and its curl is not an observable field. Its role
is a different one. For any static frame with $e^{\hat 0}=\vf\,\mathrm{d}t$ and
time independent spatial legs $e_{\hat a}^{\ i}$ ($\hat a, i=1,\dots,n$),
the trace reduces to a divergence,
\begin{equation}
a_{\hat a}=\frac{1}{\sqrt{g}}\,\partial_{i}\!\left(\sqrt{g}\,
e_{\hat a}^{\ i}\right),
\label{eq:adiv}
\end{equation}
with $g$ the determinant of the spatial metric. A nonuniform lapse adds the
gradient of its logarithm, as section~\ref{sec:wire} shows. The substitution
$\Psi=g^{-1/4}\chi$, which makes the measure flat, then brings the spatial
part of the operator to the symmetrized form
\begin{equation}
e_{\hat a}^{\ i}\gamma^{\hat a}\left(\partial_{i}+\Omega_{i}\right)\Psi
=g^{-1/4}\,\gamma^{\hat a}\,\frac{1}{2}
\left\{e_{\hat a}^{\ i},\partial_{i}\right\}\chi .
\label{eq:symmetrized}
\end{equation}
The trace term is therefore exactly the correction that renders the Dirac
operator Hermitian when the velocity tensor depends on position, the
same structure obtained from the lattice by de Juan, Sturla and Vozmediano
\cite{deJuan2012,deJuan2013}. We keep the notation $\mathcal{A}_{\mu}$ for
it, and refer to it as the geometric trace connection, but the reader should
not read it as a gauge potential.
\end{revised}

The third background is the term carrying $b^{\hat\sigma}$, and its
character depends on the number of spatial dimensions in a way that deserves
to be spelled out case by case rather than asserted in general.

\subsection{One spatial dimension}
\label{sec:n1}

On a line the frame indices run over \rev{two} values, so the totally
antisymmetric contraction in (\ref{eq:ab}) has nowhere to live and the dual
structure collapses. With the representation $\gamma^{0}=\sigma^{z}$ and
$\gamma^{1}=i\sigma^{x}$, \rev{the chirality matrix is
$\gamma^{5}=\gamma^{0}\gamma^{1}=-\sigma^{y}$, which squares to the identity.}
An explicit check reported in the appendix shows that the \rev{two}
candidate forms of $\gamma^{\hat\alpha}\Sigma^{\hat\beta\hat\gamma}$ agree
term by term,
\begin{equation}
\frac{1}{2}\left(\eta^{\hat\alpha\hat\beta}\gamma^{\hat\gamma}
-\eta^{\hat\alpha\hat\gamma}\gamma^{\hat\beta}\right)
=\rev{+}\frac{1}{2}\,\varepsilon^{\hat\beta\hat\gamma}
\gamma^{\hat\alpha}\gamma^{5}.
\label{eq:dual1d}
\end{equation}
The reason is elementary and worth stating, since it colors the
interpretation of the result. Multiplication by $\gamma^{5}$ in \rev{two
spacetime dimensions} exchanges the \rev{two} gamma matrices, as
\rev{$\gamma^{0}\gamma^{5}=\gamma^{1}$ and $\gamma^{1}\gamma^{5}=\gamma^{0}$}.
Vector and axial couplings are thus \rev{two} names for a single object,
\rev{a fact closely tied to the chiral anomaly of two dimensional models}.

What survives is therefore \rev{the trace term} and nothing else,
\begin{equation}
Q=\frac{1}{2}a_{\hat\gamma}\gamma^{\hat\gamma},
\qquad
\mathcal{A}_{\mu}=-\frac{\hbar}{2e}e^{\hat\gamma}_{\ \mu}a_{\hat\gamma},
\qquad
\mathcal{V}=0.
\label{eq:dict1d}
\end{equation}
\rev{One dimensional} Dirac systems cannot generate a scalar background out
of geometry alone. Whatever the deformation of the line, the fermion sees
\rev{at most the trace term}, and by (\ref{eq:dual1d}) the freedom to call
that term vector or axial is a matter of convention rather than a physical
distinction. \rev{Section~\ref{sec:wire} shows that a nonuniform velocity produces no
connection at all, and that the content of a one dimensional geometry lies in
the lapse.}

\subsection{Two spatial dimensions}
\label{sec:n2}

Surfaces are the case of direct experimental interest and also the one where
the dual structure is richest, because in \rev{three} spacetime dimensions
the Levi-Civita symbol has exactly as many indices as the spin connection.
Taking $\gamma^{0}=\sigma^{z}$, $\gamma^{1}=i\sigma^{x}$ and
$\gamma^{2}=i\sigma^{y}$, the spinor remains \rev{two component} and the
Clifford algebra closes on the 2D identity,
\begin{equation}
\left\{\gamma^{\hat\alpha},\gamma^{\hat\beta}\right\}
=2\eta^{\hat\alpha\hat\beta}I_{2\times2},
\qquad
\left[\gamma^{\hat\alpha},\gamma^{\hat\beta}\right]
=-2i\varepsilon^{\hat\alpha\hat\beta\hat\gamma}\gamma_{\hat\gamma}.
\label{eq:clifford2d}
\end{equation}
The product that governs the dictionary becomes
\begin{equation}
\gamma^{\hat\alpha}\Sigma^{\hat\beta\hat\gamma}
=\frac{1}{2}\left(\eta^{\hat\alpha\hat\beta}\gamma^{\hat\gamma}
-\eta^{\hat\alpha\hat\gamma}\gamma^{\hat\beta}\right)
-\frac{i}{2}\,\varepsilon^{\hat\alpha\hat\beta\hat\gamma}I_{2\times2},
\label{eq:master2d}
\end{equation}
so the dual piece is no longer accompanied by a gamma matrix. It is a
multiple of the identity in spinor space. Writing
$b\equiv\varepsilon^{\hat\alpha\hat\beta\hat\gamma}
\omega_{\hat\alpha\hat\beta\hat\gamma}$ for the single independent totally
antisymmetric contraction, equation (\ref{eq:Q}) gives
\begin{equation}
Q=\frac{1}{2}a_{\hat\gamma}\gamma^{\hat\gamma}
-\frac{i}{4}\,b\,I_{2\times2},
\label{eq:Q2d}
\end{equation}
and the curved equation reduces to
\begin{equation}
\left(\rev{\vf}\,e_{\hat\alpha}^{\ \mu}\gamma^{\hat\alpha}p_{\mu}
-\frac{i\hbar \rev{\vf}}{2}a_{\hat\gamma}\gamma^{\hat\gamma}
-\frac{\hbar \rev{\vf}}{4}\,b+m\rev{\vf^{2}}\right)\Psi=0.
\label{eq:dirac2d}
\end{equation}
Two spatial dimensions is the only case in which the triad of effective
backgrounds announced in equation (\ref{eq:model}) is realized with a
\rev{spinor scalar} in the third slot,
\begin{equation}
\rev{S_{\hat\alpha}^{\ \mu}=e_{\hat\alpha}^{\ \mu}},
\qquad
\mathcal{A}_{\mu}=-\frac{\hbar}{2e}e^{\hat\gamma}_{\ \mu}a_{\hat\gamma},
\qquad
\mathcal{V}=-\frac{\hbar \rev{\vf}}{4}\,b.
\label{eq:dict2d}
\end{equation}
\begin{revised}
The position of $\mathcal{V}$ in equation (\ref{eq:dirac2d}) shows what kind
of object it is. It sits next to $m\vf^{2}$, so it shifts the Dirac mass
locally, $m\vf^{2}\to m\vf^{2}-\hbar\vf b/4$, rather than adding a potential
energy to the Hamiltonian. With two component spinors in $2+1$ dimensions the
mass term is odd under parity and under time reversal, and $b$, which
carries an uncontracted Levi-Civita symbol, shares those properties; this is
the sense in which we call it a pseudoscalar. Because the three frame indices
must be distinct, one of them is the time label, and $b$ vanishes for every
static frame in which $e^{\hat 0}=\vf\,\mathrm{d}t$ and the spatial legs do
not depend on time. Section~\ref{sec:stationary} shows that frames which mix
time and space, such as those of a rotating or dragged medium, do produce it.
\end{revised}

\subsection{Three spatial dimensions}
\label{sec:n3}

In \rev{four} spacetime dimensions the Weyl representation is the natural
choice,
\begin{equation}
\gamma^{\hat\mu}=\left[\begin{array}{cc}
0 & \sigma^{\hat\mu}\\
\bar{\sigma}^{\hat\mu} & 0
\end{array}\right],
\quad
\sigma^{\hat\mu}=\left(I_{2\times2},\sigma^{x},\sigma^{y},\sigma^{z}\right),
\quad
\bar{\sigma}^{\hat\mu}=\left(I_{2\times2},-\sigma^{x},-\sigma^{y},
-\sigma^{z}\right),
\label{eq:weyl}
\end{equation}
for which $\gamma^{5}=i\gamma^{0}\gamma^{1}\gamma^{2}\gamma^{3}
=\mathrm{diag}(-I_{2\times2},I_{2\times2})$. This is the case treated by
equations (\ref{eq:master}) to (\ref{eq:dirac4d}), and the dictionary reads
\begin{equation}
\rev{S_{\hat\alpha}^{\ \mu}=e_{\hat\alpha}^{\ \mu}},
\qquad
\mathcal{A}_{\mu}=-\frac{\hbar}{2e}e^{\hat\gamma}_{\ \mu}a_{\hat\gamma},
\qquad
\mathcal{V}=-\frac{\hbar \rev{\vf}}{4}\,b^{\hat\sigma}\gamma_{\hat\sigma}\gamma^{5}.
\label{eq:dict3d}
\end{equation}
The third entry is no longer a scalar. It carries a free frame index and a
$\gamma^{5}$, which makes it an axial vector potential coupling to the chiral
current rather than a \rev{mass term}. Calling it a scalar background, as the
2D intuition invites one to do, would misrepresent both its transformation
properties and its physical effect.

\begin{revised}
The two kinds of component act differently, and both have a clear meaning
once the equation is written in Hamiltonian form. The time component enters
the Hamiltonian as a term proportional to $(\hbar\vf/4)\,b^{\hat 0}\gamma^{5}$, a chiral chemical potential that
shifts the nodes of opposite chirality in opposite directions in energy. The
spatial components enter as a term proportional to $(\hbar\vf/4)\,b^{\hat k}\,\mathsf{S}^{k}$,
with $\mathsf{S}^{k}\equiv\gamma^{5}\gamma^{0}\gamma^{\hat k}$ the spin matrix, and shift the nodes of
opposite chirality in opposite directions in momentum. In 3D Dirac and Weyl
materials \cite{Armitage2018} these are the two familiar axial backgrounds,
an energy separation and a momentum separation of the nodes, and neither is
a mass. Whether a given $b^{\hat\sigma}$ is physical, however, depends on the
frame, as section~\ref{sec:frames} explains and section~\ref{sec:three}
illustrates with a twisted frame in flat space.
\end{revised}

The pattern across the \rev{three} cases is set by how much room the
Levi-Civita symbol has. On a line the dual contraction degenerates into the
trace, on a surface it saturates the spinor indices and produces a
\rev{pseudoscalar mass}, and in a volume it leaves one index free and
produces an axial vector. The trace contraction, by contrast, always yields
the term of equation (\ref{eq:Amu}). \rev{That term is the only entry
of the dictionary that can appear in every dimension, but it is not the pseudogauge field that experiments on strained
samples detect \cite{Levy2010,Jiang2017}; that field has a lattice origin, as
section~\ref{sec:scope} discusses.}

\subsection{Behavior under local frame rotations}
\label{sec:frames}

\begin{revised}
The orthonormal frame is defined only up to a local Lorentz transformation
$e^{\hat\alpha}_{\ \mu}\to\Lambda^{\hat\alpha}_{\ \hat\beta}(x)\,
e^{\hat\beta}_{\ \mu}$. The Dirac operator is covariant under it, provided
the spinor transforms with the corresponding element of the spin group, but
the spin connection picks up an inhomogeneous term,
$\omega\to\Lambda\omega\Lambda^{-1}+\Lambda\,\mathrm{d}\Lambda^{-1}$. That
term feeds into both contractions of (\ref{eq:ab}), so $a_{\hat\gamma}$,
$b^{\hat\sigma}$, $\mathcal{A}_{\mu}$ and its curl all change with the frame.
The correspondence (\ref{eq:model}) holds in every frame, but its entries are
frame dependent, and only combinations that are insensitive to $\Lambda$ can
be given a physical meaning.

On a static surface the relevant transformations are rotations of the
spatial legs by an angle $\theta(x,y)$,
$e^{\hat 1}\to\cos\theta\,e^{\hat 1}+\sin\theta\,e^{\hat 2}$ and
$e^{\hat 2}\to-\sin\theta\,e^{\hat 1}+\cos\theta\,e^{\hat 2}$. A direct
computation gives
\begin{equation}
\delta\mathcal{A}_{i}=-\frac{\hbar}{2e}\,\sqrt{g}\,
\varepsilon_{ij}\,g^{jk}\partial_{k}\theta,
\qquad
\delta\mathcal{F}_{xy}=\frac{\hbar}{2e}\,\sqrt{g}\,\Delta_{g}\theta,
\label{eq:rotation}
\end{equation}
with $\varepsilon_{xy}=1$, $\mathcal{F}_{xy}=\partial_{x}\mathcal{A}_{y}
-\partial_{y}\mathcal{A}_{x}$, $g_{ij}$ the spatial metric and $\Delta_{g}$
its Laplace-Beltrami operator. In a flat plane with a rotated frame this is
the result $\mathcal{F}_{xy}=(\hbar/2e)\nabla^{2}\theta$, which is not zero
although the plane is flat. The divergence, by contrast, does not change,
and it carries the curvature,
\begin{equation}
\nabla_{i}\mathcal{A}^{i}=\frac{\hbar}{2e}\,K,
\label{eq:divK}
\end{equation}
where indices are raised with $g^{ij}$ and $K$ is the Gaussian curvature of
the surface. Equations (\ref{eq:rotation}) and (\ref{eq:divK}) say that the
trace connection on a surface is, up to a factor, the Hodge dual of the
spin connection. Its curl is pure frame choice, and its divergence is
geometry.

In 2+1 dimensions the pseudoscalar $b$ is also affected by rotations that
depend on time, and in 3+1 dimensions $b^{\hat\sigma}$ is affected by
rotations that depend on position. Section~\ref{sec:three} gives an
explicit case in which a frame twisted along one axis produces a constant
$b^{\hat 0}$ in flat space.
\end{revised}

% ---------------------------------------------------------------------------
\section{Applications}
\label{sec:applications}

\subsection{Membranes and the choice of frame}
\label{sec:membrane}
Two spatial dimensions is where the dictionary meets the laboratory, so it
deserves an explicit deformation rather than a statement about a general
metric. A static membrane is described by the frame
\begin{equation}
e^{\hat 0}=\rev{\vf}\,\mathrm{d}t,
\qquad
e^{\hat 1}=\lambda_{1}(x,y)\,\mathrm{d}x,
\qquad
e^{\hat 2}=\lambda_{2}(x,y)\,\mathrm{d}y,
\label{eq:frame2d}
\end{equation}
with positive functions $\lambda_{1}$ and $\lambda_{2}$ that encode how much
the sample is stretched along each direction. Building the Christoffel
symbols, feeding them into (\ref{eq:spinconn}), projecting with
(\ref{eq:ricci}) and contracting as in (\ref{eq:ab}) is a mechanical but
lengthy computation whose outcome is compact,
\begin{equation}
\mathcal{A}_{\mu}\mathrm{d}x^{\mu}
=-\frac{\hbar}{2e}\left(\partial_{x}\ln\lambda_{2}\,\mathrm{d}x
+\partial_{y}\ln\lambda_{1}\,\mathrm{d}y\right),
\qquad
b=0.
\label{eq:A2d}
\end{equation}
The crossing of labels is not a misprint. The \rev{component} along $x$ is
governed by the stretching along $y$ and the other way round, which is
already a hint that isotropic and anisotropic deformations will behave
differently.

Consider first a conformal profile, $\lambda_{1}=\lambda_{2}=e^{\varphi}$.
Equation (\ref{eq:A2d}) collapses to
$\mathcal{A}_{\mu}=-(\hbar/2e)\partial_{\mu}\varphi$, a pure gradient.
Because of the factor of $i$ that accompanies the minimal coupling in
(\ref{eq:model}), removing a gradient is not a change of phase but a change
of normalization. \rev{The substitution $\Psi=e^{-\varphi/2}\chi$ eliminates
the connection term, but it leaves the velocity and the measure untouched.
Writing $H_{0}=-i\hbar\vf\,\gamma^{0}\gamma^{\hat a}\partial_{a}$ for the flat
massless operator, the spatial part of the equation becomes
\begin{equation}
-i\hbar\vf\,e^{-\varphi}\gamma^{0}\gamma^{\hat a}
\left(\partial_{a}+\Omega_{a}\right)\Psi
=e^{-\varphi/2}\left(e^{-\varphi}H_{0}\right)\chi,
\label{eq:conformal}
\end{equation}
while the norm becomes $\int\sqrt{g}\,|\Psi|^{2}\,\mathrm{d}^{2}x
=\int e^{\varphi}|\chi|^{2}\,\mathrm{d}^{2}x$. The problem for $\chi$ is a Dirac operator with the isotropic, position
dependent velocity $\vf e^{-\varphi}$ and a nonuniform measure. Equivalently,
in terms of $\chi'=e^{\varphi/2}\chi$ the operator takes the symmetrized form
(\ref{eq:symmetrized}) with flat measure. The geometric background of a
conformally deformed membrane is therefore a modulation of the Fermi velocity
and nothing else.} This is the same
structure exploited in the program that turns negatively curved graphene into
a table top realization of quantum field theory in curved space
\cite{Iorio2011,IorioLambiase2014,Gallerati2021}.

\begin{figure}[htb]
\centering
\includegraphics[width=0.95\textwidth]{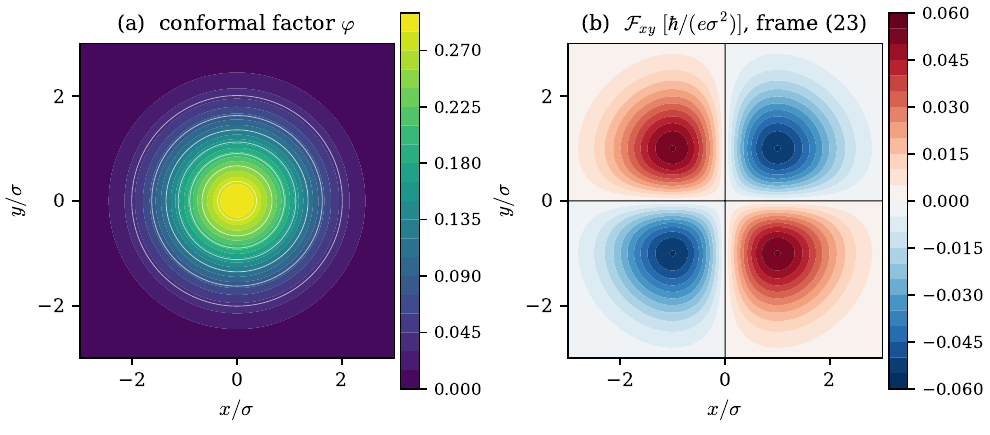}
\caption{\rev{Conformal factor $\varphi=\varepsilon\exp[-(x^{2}+y^{2})/
(2\sigma^{2})]$ with $\varepsilon=0.3$. Panel (a) shows $\varphi$, which
enters the frame (\ref{eq:frame2d}); it is a property of the frame and not
a height profile. Panel (b) shows the curl (\ref{eq:curl2d}) for the
anisotropic choice $\lambda_{1}=e^{\varphi}$, $\lambda_{2}=1$,
$\mathcal{F}_{xy}=-(\hbar/2e)\,\partial_{x}\partial_{y}\varphi$, in units of
$\hbar/(e\sigma^{2})$; its maximum modulus is $\varepsilon/(2\mathrm{e})$,
with $\mathrm{e}$ the base of the natural logarithm. The quadrupolar pattern follows the mixed
derivative of $\varphi$. For the isotropic choice
$\lambda_{1}=\lambda_{2}=e^{\varphi}$ the curl vanishes at every point.
Both statements refer to the diagonal frame; by equation (\ref{eq:rotation})
a local rotation of the frame changes panel (b).}}
\label{fig:pseudo}
\end{figure}

Whether the \rev{connection term can be removed by a rescaling of the spinor
in the frame (\ref{eq:frame2d})} is decided by the curl of (\ref{eq:A2d}).
Writing $\mathcal{F}_{ij}=\partial_{i}\mathcal{A}_{j}
-\partial_{j}\mathcal{A}_{i}$ and inserting the potential gives the closed
form
\begin{equation}
\rev{\mathcal{F}_{ij}
=-\frac{\hbar}{2e}\,\varepsilon_{ij}\,\partial_{x}\partial_{y}
\ln\!\left(\frac{\lambda_{1}}{\lambda_{2}}\right)},
\label{eq:curl2d}
\end{equation}
\rev{with $\varepsilon_{xy}=-\varepsilon_{yx}=1$. Written in this way the
antisymmetry in $ij$ is explicit. The mixed derivative is symmetric, but the
logarithm is odd under the relabeling $x\leftrightarrow y$, which also
exchanges $\lambda_{1}$ and $\lambda_{2}$.}
The curl vanishes if and only if the ratio $\lambda_{1}/\lambda_{2}$
factorizes into a function of $x$ times a function of $y$. \rev{Within the
diagonal frame (\ref{eq:frame2d}) and the laboratory coordinates in which it
is written, isotropy is a sufficient condition and anisotropy that varies
along both directions obstructs the rescaling.

This statement must not be read as a property of the membrane. By equation
(\ref{eq:rotation}), a rotation of the frame with
$\sqrt{g}\,\Delta_{g}\theta=-(2e/\hbar)\mathcal{F}_{xy}$ removes the curl,
and such an angle always exists locally. In the rotated frame the trace
connection becomes a gradient, $\mathcal{A}=-(\hbar/2e)\,\mathrm{d}f$, and
equation (\ref{eq:divK}) fixes the function through
\begin{equation}
\Delta_{g}f=-K .
\label{eq:poisson}
\end{equation}
The same conclusion follows from the existence of isothermal coordinates, in
which every 2D metric takes the conformal form \cite{Chern1955}: there
$\lambda_{1}=\lambda_{2}$, the curl vanishes, and $f=\varphi$ solves
(\ref{eq:poisson}). For a static surface, after a suitable rotation of the
frame, the connection term can always be traded for a rescaling of the
spinor whose logarithm is sourced by the Gaussian curvature, and the geometry reaches the spectrum only through the
velocity tensor and the measure.} Figure~\ref{fig:pseudo} \rev{shows the
curl (\ref{eq:curl2d}) for a smooth Gaussian conformal factor in the diagonal
frame with $\lambda_{1}=e^{\varphi}$, $\lambda_{2}=1$, and its vanishing for
$\lambda_{1}=\lambda_{2}$.}

Shear behaves in the same spirit but with a different anatomy. Replacing the
orthogonal frame by $e^{\hat 1}=\mathrm{d}x+u(x,y)\,\mathrm{d}y$ and
$e^{\hat 2}=\mathrm{d}y$ leaves a single nonvanishing component,
$\mathcal{A}_{y}=(\hbar/2e)\partial_{x}u$, and \rev{a curl}
$\mathcal{F}_{xy}=(\hbar/2e)\partial_{x}^{2}u$ controlled by the curvature of
the shear profile along the direction it displaces. A surface of revolution,
by contrast, with $\lambda_{1}=1$ and $\lambda_{2}=f(x)$, gives
$\mathcal{A}_{x}=-(\hbar/2e)\partial_{x}\ln f$ and no \rev{curl},
as equation (\ref{eq:curl2d}) anticipates.

\subsection{A Gaussian bump}
\label{sec:bump}

\begin{figure}[htb]
\centering
\includegraphics[width=\textwidth]{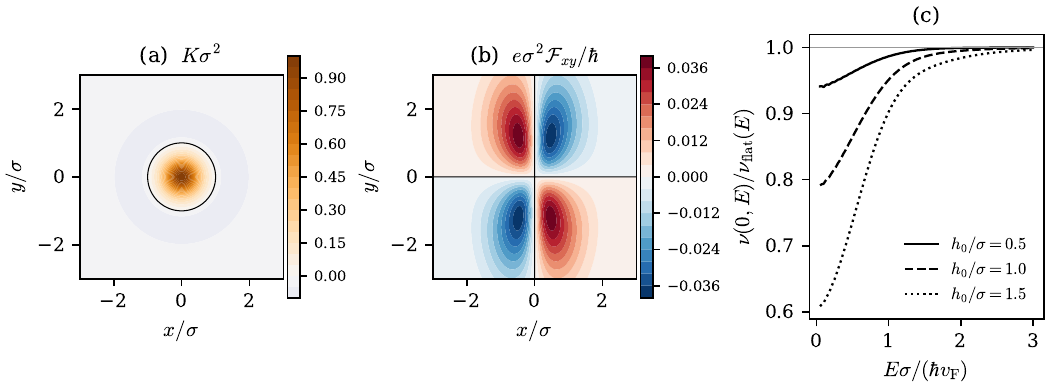}
\caption{\rev{Gaussian bump (\ref{eq:bump}). (a) Gaussian curvature
$K\sigma^{2}$ for $h_{0}=\sigma$; the black line marks $K=0$. (b) Curl
$\mathcal{F}_{xy}$ of the trace connection, in units of
$\hbar/(e\sigma^{2})$, in the symmetric frame for an elliptical bump with
widths $\sigma$ along $x$ and $2\sigma$ along $y$ and $h_{0}=\sigma$; for the
circular bump the same quantity vanishes identically. (c) Local density of
states at the center of the circular bump divided by its flat value, from the
radial problem (\ref{eq:radial}); the low energy limit is given by
equation (\ref{eq:ldoslow}). Only the geometric contribution is included.}}
\label{fig:bump}
\end{figure}

\begin{revised}
The conformal factor of the previous section is a property of a frame. A
deformed sample is described instead by its height $h(x,y)$ above the
reference plane, and the metric it induces on the carriers is
$g_{ij}=\delta_{ij}+\partial_{i}h\,\partial_{j}h$. We take the Gaussian bump
\begin{equation}
h(x,y)=h_{0}\exp\!\left(-\frac{x^{2}+y^{2}}{2\sigma^{2}}\right),
\label{eq:bump}
\end{equation}
which is the standard idealization of the nanobubbles and corrugations of the
experiments cited in the introduction, and whose Gaussian curvature at the
apex is $K(0)=h_{0}^{2}/\sigma^{4}$. Panel (a) of Figure~\ref{fig:bump} shows
$K$ for $h_{0}=\sigma$: positive on the cap, weakly negative on the
surrounding ring where the bump meets the plane.

Two frames are natural here. The symmetric frame $e^{\hat a}_{\ i}
=(g^{1/2})_{ai}$, built from the matrix square root of $g_{ij}$, stays
close to the Cartesian axes. The polar frame $e^{\hat 1}=\lambda(r)\,
\mathrm{d}r$, $e^{\hat 2}=r\,\mathrm{d}\phi$, with
$\lambda=[1+h'(r)^{2}]^{1/2}$, is diagonal in polar coordinates and gives
$\mathcal{F}_{r\phi}=0$ by equation (\ref{eq:curl2d}), since
$\lambda_{1}/\lambda_{2}=\lambda(r)/r$ does not depend on $\phi$. For the
circular bump the symmetric frame also gives $\mathcal{F}_{xy}=0$ at every
point, because it differs from the polar frame by a rotation through the
polar angle, which is harmonic away from the apex. This is in line with the
perturbative finding of de Juan, Cortijo and Vozmediano that the geometric
gauge term does not contribute at first order for axially symmetric bumps
\cite{deJuan2007}. An elliptical bump with widths $\sigma$ and $2\sigma$
breaks that coincidence; panel (b) shows the resulting curl in the symmetric
frame, a quadrupole of maximum modulus $0.039\,\hbar/(e\sigma^{2})$ for
$h_{0}=\sigma$, which a further rotation of the frame would remove.

The frame independent content of the bump is in the spectrum. For the
circular bump the problem separates in the polar frame. Writing the proper
radial distance as $\rho$ and the radius of the circle of constant $\rho$ as
$R(\rho)=r$, the spinor $\sqrt{R}\,\Psi$, with components referred to the
polar frame, obeys in each channel of total angular
momentum $j\in\mathbb{Z}+1/2$,
\begin{equation}
H_{j}=\hbar\vf\left(-i\sigma^{x}\partial_{\rho}+\sigma^{y}\,
\frac{j}{R(\rho)}\right),
\label{eq:radial}
\end{equation}
which reduces to the flat problem when $R=\rho$. Only the channels
$j=\pm1/2$ reach the center. Integrating the regular solution outwards and
comparing its asymptotic amplitude with the flat one gives the local density
of states at the apex relative to its flat value, without a finite box or an
artificial broadening. Panel (c) of Figure~\ref{fig:bump} shows the result.
The density is reduced at every energy, and the reduction disappears once the
wavelength is shorter than the width of the bump. At low energy, where the
wavelength exceeds the bump, the regular solution is amplified across the cap
by a factor that can be computed in closed form, and
\begin{equation}
\frac{\nu(0,E\to0)}{\nu_{\mathrm{flat}}(E)}
=\exp\!\left[-\int_{0}^{\infty}\bigl(\lambda(r)-1\bigr)\frac{\mathrm{d}r}{r}
\right],
\label{eq:ldoslow}
\end{equation}
which gives $0.940$, $0.790$ and $0.605$ for $h_{0}/\sigma=0.5$, $1$ and
$1.5$ and reproduces the numerical curves at the lowest energies. The sign
agrees with the perturbative result of Ref.~\cite{deJuan2007}, which
found the corrected density always below the flat one. The calculation keeps
only the geometric part of the problem; the lattice pseudogauge field of a
real bubble, which has a threefold pattern and produces pseudo Landau levels
\cite{Levy2010,Vozmediano2010}, has to be added on top of it.
\end{revised}

\subsection{One spatial dimension: velocity and lapse}
\label{sec:wire}

\begin{figure}[htb]
\centering
\includegraphics[width=0.95\textwidth]{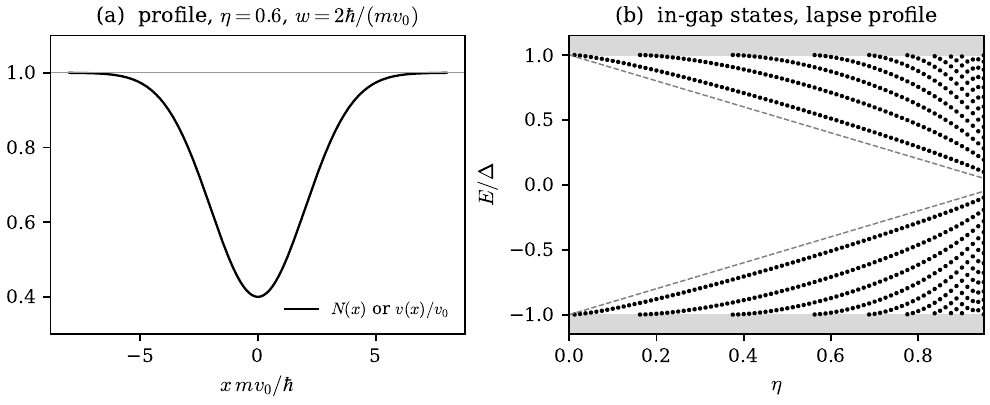}
\caption{\rev{A massive Dirac fermion on a line. (a) The profile
$1-\eta\exp[-x^{2}/(2w^{2})]$ with $\eta=0.6$ and $w=2\hbar/(mv_{0})$, used
either as a velocity $v(x)/v_{0}$ or as a lapse $N(x)$. (b) In-gap energies
for the lapse, from equation (\ref{eq:lapse}), as a function of $\eta$. The
shaded bands are the continuum and the dashed lines mark $\pm(1-\eta)\Delta$.
A velocity profile alone produces no in-gap states.}}
\label{fig:wire}
\end{figure}

\begin{revised}
A line has no intrinsic curvature, so the interest of the 1D case lies in
how the frame is chosen in time and space. Consider first a static frame
$e^{\hat 0}=v_{0}\,\mathrm{d}t$, $e^{\hat 1}=v_{0}\,\mathrm{d}x/v(x)$ with
$v(x)>0$, which describes a wire with a nonuniform Fermi velocity
$v(x)$. The spin connection vanishes identically: in one spatial dimension
there is no rotation of a static frame. The Dirac operator is
$-i\hbar\,\gamma^{0}\gamma^{1}v(x)\partial_{x}$, which is Hermitian with
respect to the proper measure $v_{0}\,\mathrm{d}x/v$, and the coordinate
$s=\int\mathrm{d}x\,v_{0}/v(x)$ maps it onto a flat operator with constant
mass. The spectrum of $H^{2}$ is therefore bounded below by the square of
the gap $\Delta=mv_{0}^{2}$, and no modulation of the velocity alone, however
strong, can bind a state inside the gap. What remains of the velocity profile
is a reparametrization of the wire, whose only global trace is the total time
of flight $\int\mathrm{d}x/v$.

The lapse behaves differently. For $e^{\hat 0}=N(x)\,v_{0}\,\mathrm{d}t$ and
$e^{\hat 1}=\mathrm{d}x$ the trace connection is the gradient
$\mathcal{A}_{x}=-(\hbar/2e)\,\partial_{x}\ln N$, removed by
$\Psi=N^{-1/2}\chi$. Multiplying the equation by $N$, dividing by $v_{0}$ and passing to
$\mathrm{d}s=\mathrm{d}x/N$ gives, with $x^{0}=v_{0}t$,
\begin{equation}
\left(-i\hbar\gamma^{0}\partial_{0}-i\hbar\gamma^{1}\partial_{s}
+mv_{0}\,N(s)\right)\chi=0,
\label{eq:lapse}
\end{equation}
a flat Dirac equation with a position dependent mass $N(s)\Delta$. A lapse
modulates the velocity and the gap together, keeping their ratio fixed, and
it is that joint modulation that carries the geometric content of a 1D
background. Physically, it describes a 1D Dirac system in which the overall
energy scale of the hopping varies along the wire, for instance a dimerized
chain in which the mean hopping and its dimerization are rescaled by the same
factor $N(x)$, so that velocity and gap vary together.
%\edit{Note for David: please confirm, or replace with a concrete
%realization you prefer, the physical interpretation of the lapse given in the
%previous sentence. We did not find a reference that states it in these terms.}

Equation (\ref{eq:lapse}) is the setting in which the factorization methods
mentioned in the introduction apply directly. With $W(s)=N(s)\,mv_{0}/\hbar$,
the square of the Hamiltonian splits into the supersymmetric partners
$-\partial_{s}^{2}+W^{2}\mp W'$ \cite{Cooper1995}, and a dip in $N$ binds
states inside the gap. Figure~\ref{fig:wire} shows the in-gap spectrum for
$N(x)=1-\eta\exp[-x^{2}/(2w^{2})]$ with $w=2\hbar/(mv_{0})$. One bound state
pair appears as soon as $\eta>0$, and more follow as the well deepens; for
$\eta=0.6$ the bound energies are $|E|/\Delta=0.510$, $0.710$, $0.872$ and
$0.985$. The lowest level approaches $(1-\eta)\Delta$, the minimum of the
local gap, as the well grows. Since $W>0$ everywhere, no zero mode appears
and the spectrum is symmetric. For the velocity-only profile with the same
$\eta$ the same computation finds no state inside the gap, as it must.
\end{revised}

\subsection{Stationary frames on a surface}
\label{sec:stationary}

\begin{figure}[htb]
\centering
\includegraphics[width=0.95\textwidth]{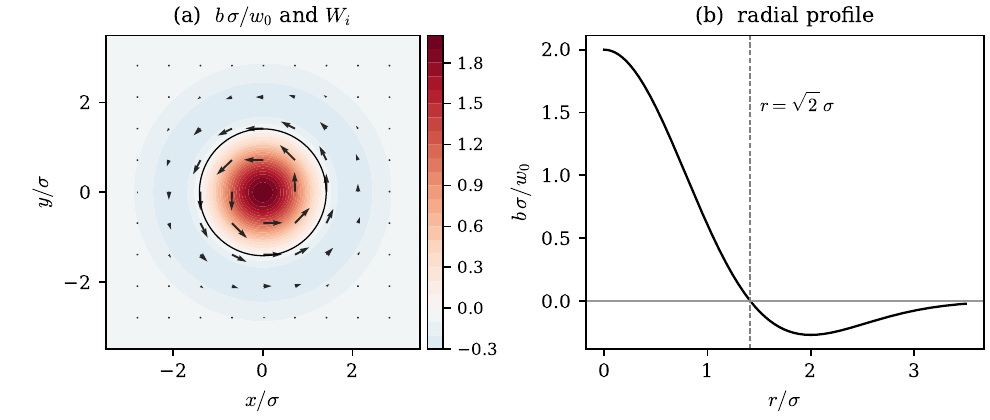}
\caption{\rev{Pseudoscalar $b$ for the circulating drag of equation
(\ref{eq:bvortex}). (a) Map of $b$ in units of $w_{0}/\sigma$, with arrows
indicating $W_{i}$ and the black line marking $b=0$. (b) Radial profile; the
sign changes at $r=\sqrt{2}\,\sigma$. By equation (\ref{eq:dict2d}) the
local Dirac mass is shifted by $-\hbar\vf b/4$.}}
\label{fig:drag}
\end{figure}

\begin{revised}
A frame whose time leg has spatial components describes a medium that drags
the carriers, the analogue of a gravitomagnetic background. For
\begin{equation}
e^{\hat 0}=\mathrm{d}x^{0}+W_{i}(x,y)\,\mathrm{d}x^{i},
\qquad
e^{\hat 1}=\mathrm{d}x,
\qquad
e^{\hat 2}=\mathrm{d}y,
\label{eq:drag}
\end{equation}
with $x^{0}=\vf t$ and dimensionless $W_{i}$, the trace vanishes and the
pseudoscalar is the curl of the drag,
\begin{equation}
a_{\hat\gamma}=0,
\qquad
b=\partial_{x}W_{y}-\partial_{y}W_{x}.
\label{eq:bdrag}
\end{equation}
%The case $W=W(x,y)\,\mathrm{d}y$ considered in the first version of this
%manuscript is the special case $b=\partial_{x}W$. 
By equation
(\ref{eq:dict2d}) the drag shifts the Dirac mass locally by
$-\hbar\vf b/4$. For a localized circulating drag of strength $w_{0}$ and width $\sigma$,
$W_{x}=-w_{0}\,(y/\sigma)\,e^{-r^{2}/(2\sigma^{2})}$ and
$W_{y}=w_{0}\,(x/\sigma)\,e^{-r^{2}/(2\sigma^{2})}$, the pseudoscalar is
\begin{equation}
b=\frac{w_{0}}{\sigma}\,e^{-r^{2}/(2\sigma^{2})}
\left(2-\frac{r^{2}}{\sigma^{2}}\right),
\label{eq:bvortex}
\end{equation}
positive in the core and negative on the surrounding ring, with a change of
sign at $r=\sqrt{2}\,\sigma$, as figure~\ref{fig:drag} shows. The local mass changes sign near the core
when $m\vf^{2}<\hbar\vf w_{0}/(2\sigma)$. Since the surfaces of constant $t$
must remain spacelike, which requires $|W|<1$ and hence
$w_{0}<\mathrm{e}^{1/2}$, a sign change is possible only for
$\sigma\lesssim\hbar/(m\vf)$. The region of inverted mass is then no larger
than the Compton length, and the familiar picture of states bound to a mass
domain wall cannot be taken over without a separate calculation.

A rigidly rotating frame belongs to the same family. In coordinates that
rotate with angular velocity $\Omega$, $e^{\hat 1}=\mathrm{d}x+(\Omega/\vf)\,
y\,\mathrm{d}x^{0}$ and $e^{\hat 2}=\mathrm{d}y-(\Omega/\vf)\,x\,
\mathrm{d}x^{0}$, one finds $a_{\hat\gamma}=0$ and the uniform value
$b=-2\Omega/\vf$. The mass shifts by $\hbar\Omega/2$. This uniform $b$ is
physical only because the material itself rotates with the frame, and together with the orbital term
$-\Omega L_{z}$ the shift completes the coupling $-\Omega J_{z}$ of a spin one
half particle to rotation \cite{Mashhoon1988,HehlNi1990}. In graphene the two
spinor components refer to the sublattice pseudospin rather than to the
electron spin, so the identification presumes that the pseudospin carries
angular momentum, as argued by Mecklenburg and Regan \cite{Mecklenburg2011}. The
size of the effect deserves a plain statement: for a mechanical rotation at
$\Omega/2\pi=10$\,kHz the shift is of order $10^{-11}$\,eV. The point of the
example is structural. A static bend of a sample produces no pseudoscalar at
all, and any genuine $b$ on a surface signals a frame that mixes time and
space or, as discussed in section~\ref{sec:scope}, a connection with torsion.
\end{revised}

\subsection{Twisted and rotating frames in a volume}
\label{sec:three}

\begin{revised}
Two frames in flat 3D space make the frame dependence of the axial term
explicit and separate its artificial and physical parts. In the first, the
spatial legs rotate about the $z$ axis as one moves along it,
\begin{equation}
e^{\hat 1}=\cos(\tau z)\,\mathrm{d}x+\sin(\tau z)\,\mathrm{d}y,
\qquad
e^{\hat 2}=-\sin(\tau z)\,\mathrm{d}x+\cos(\tau z)\,\mathrm{d}y,
\qquad
e^{\hat 3}=\mathrm{d}z,
\label{eq:twist}
\end{equation}
with $e^{\hat 0}=\mathrm{d}x^{0}$. The space is flat, yet
$a_{\hat\gamma}=0$ and $b^{\hat\sigma}=(2\tau,0,0,0)$, a term with the form of a uniform chiral
chemical potential of magnitude $\hbar\vf\tau/2$. It is entirely an artifact
of the frame: the spin rotation $\exp(-i\tau z\mathsf{S}^{3}/2)$ that undoes the
twist removes it. This example answers
in the negative the question of whether $b^{\hat\sigma}$ is observable by
itself. A twist of the frame is not a twist of the material, and a lattice
twist enters the Dirac operator through a different route, as an axial
pseudogauge field produced by the modulation of the node separation
\cite{Cortijo2015,Pikulin2016,Ilan2020}.

In the second frame the coordinates rotate rigidly about $z$,
$e^{\hat 1}=\mathrm{d}x+(\Omega/\vf)\,y\,\mathrm{d}x^{0}$,
$e^{\hat 2}=\mathrm{d}y-(\Omega/\vf)\,x\,\mathrm{d}x^{0}$,
$e^{\hat 3}=\mathrm{d}z$. Again $a_{\hat\gamma}=0$, and now
$b^{\hat\sigma}=(0,0,0,2\Omega/\vf)$, a spatial axial vector along the
rotation axis. In Hamiltonian form it is the spin part $\hbar\Omega\mathsf{S}^{3}/2$ of the
coupling $-\boldsymbol{\Omega}\cdot\mathbf{J}$ between angular momentum and
rotation \cite{HehlNi1990}, which for a Weyl node of chirality $\chi=\pm1$
shifts the node along $k_{z}$ by $-\chi\,\Omega/(2\vf)$, up to an overall sign
fixed by conventions. Here the frame is tied to a physical observer, the
rotating sample. The orbital part $-\Omega L_{z}$ is also present, so the
statement concerns the spin part of the coupling rather than a conserved
momentum. In a Weyl material $\mathsf{S}^{3}$ acts on the band pseudospin, and
its identification with angular momentum carries the same caveat as in
graphene. It is also the microscopic ingredient
behind the chiral vortical effect, the axial current that a rotating chiral
system carries along its axis \cite{Vilenkin1979,Landsteiner2011}. For
$\Omega/2\pi=10$\,kHz and $\vf=3\times10^{5}$\,m/s the shift is about
$0.1$\,m$^{-1}$, far below any lattice scale; the value of the example is
again that it shows which entry of the dictionary rotation feeds, and that
the answer is an axial vector rather than a mass.
\end{revised}

% ---------------------------------------------------------------------------
\section{Scope of the correspondence}
\label{sec:scope}
Nothing in sections \ref{sec:setup} and \ref{sec:dictionary} involves an
expansion. The map from curvature to effective backgrounds is an identity
between two ways of writing the same operator, valid for any invertible
vielbein, and in that sense it does not degrade as the deformation grows.
Approximations enter one step earlier, when a physical displacement field of
a real sample is translated into a frame, and one step later, when the flat
space problem is actually solved. The second point matters for geometries
such as the graphene wormhole, where the deformation near the throat is in no
sense small \cite{Gonzalez2010}: the dictionary still applies, but treating
the resulting backgrounds perturbatively does not.

A more delicate issue concerns what the \rev{trace connection} of equation
(\ref{eq:Amu}) actually is. It descends from the spin connection, \rev{it is
anti-Hermitian, and section~\ref{sec:dictionary} shows that it is the term
that keeps the operator Hermitian when the velocity depends on position; its
curl depends on the frame and is not observable.} \rev{It is not the pseudogauge field}
that dominates the pseudomagnetic response measured in strained samples. That
one has a lattice origin, arises from the modulation of the hopping
amplitudes and enters the continuum description as a valley dependent vector
potential \rev{that is Hermitian and produces pseudo Landau levels}
\cite{Vozmediano2010,Guinea2010,deJuan2013}. The \rev{two} contributions
coexist and have been reconciled explicitly within the covariant formalism
\cite{deJuan2012,Amorim2016}, \rev{where the modulation of the hopping
amplitudes produces both a correction to the velocity tensor, which can be
absorbed into the frame of equation (\ref{eq:strain}), and a valley
dependent vector potential, which has no counterpart in $\Omega_{\mu}$.} Reporting the geometric piece as if it accounted for the
fields inferred from pseudo Landau levels
\cite{Levy2010,Meng2013,Jiang2017} would therefore misattribute the effect.
\rev{What the geometry does leave in the spectrum is a modulation of the
density of states governed by the velocity tensor and, on a surface, by the
Gaussian curvature, as the bump of section~\ref{sec:bump} illustrates.}

The silence of the third background in static geometries invites a second
remark. Its source, the totally antisymmetric part of the Ricci rotation
coefficients, is exactly the quantity that a connection with torsion makes
generically nonzero \rev{\cite{HehlDatta1971,Shapiro2002}}, and torsion is
the natural language for dislocations and disclinations in a continuum
description of defective media \cite{KatanaevVolovich1992,Katanaev2005}.
Topological defects in graphene have long been treated along those lines
\cite{Cortijo2007,CortijoEPL2007}, so the appearance of a \rev{pseudoscalar
mass} in a 2D Dirac system should be read as evidence of torsion or of a
\rev{frame that mixes time and space} rather than as evidence of curvature.
The present work assumes a Levi-Civita connection throughout, and extending
the dictionary to Riemann-Cartan backgrounds is the obvious next step.

Chirality deserves a closing caution. In \rev{three} spatial dimensions the
third entry of the dictionary couples through
$\gamma^{\hat\sigma}\gamma^{5}$, so it acts with opposite sign on the
\rev{two} chiral components of the spinor. Any statement transported from the
2D case, where that entry is a single function multiplying the identity
\rev{and shifting the mass}, has to be re-examined before being applied to a
Weyl or Dirac semimetal. The same warning applies in reverse to one spatial
dimension, where equation (\ref{eq:dual1d}) makes the vector and axial
descriptions indistinguishable and no amount of algebra will separate effects
that the 2D intuition keeps apart.

% ---------------------------------------------------------------------------
\section{Concluding remarks}
\label{sec:conclusions}
Reading the geometry of a Dirac system through the Ricci rotation
coefficients turns a question about manifolds into a question about
\rev{two} contractions, and the answer to that question changes with
dimension in a way that is easy to overlook when one works exclusively on
surfaces. The practical consequence is that results migrating from graphene
to 3D Dirac and Weyl materials should be checked term by term. A
\rev{pseudoscalar shift of the Dirac mass} and an axial potential coupling to
the chiral current do different things to a spectrum, and only one of them
exists in each setting.

\rev{The applications show where the correspondence pays off and where it
needs care. Its entries depend on the frame, so the statements that survive
are about invariant combinations: on a surface, the divergence of the trace
connection, which is the Gaussian curvature, and the velocity tensor that
governs the density of states of a bump; on a line, the lapse, which
modulates gap and velocity together and can bind states that a velocity
profile cannot; on a surface or in a volume, the pseudoscalar and axial terms
of frames tied to physical observers, which reproduce the coupling between
spin and rotation. Twisted frames in flat space show how easily an axial
term can be produced by the frame alone.}

Several directions open from here. Extending the construction to
Riemann-Cartan backgrounds is the most immediate, since torsion is precisely
what switches on the contraction that stays silent for static membranes, and
dislocations and disclinations in a defective lattice are described exactly
in that language. A second direction is computational. Because the
correspondence is an identity rather than an expansion, it can be fed to
numerical scattering codes written for flat space, with the deformation
entering through the frame; geometries with a throat or a neck, where no
small parameter exists, are the natural test cases. A third is the
systematic use of factorization methods on the resulting \rev{one
dimensional} radial problems, \rev{of which equation (\ref{eq:lapse}) is the
simplest instance}.

What would sharpen the picture most is an experiment sensitive to the
distinction drawn in section \ref{sec:scope}. The lattice gauge field and the
geometric \rev{terms} enter the same Dirac operator through different doors.
\rev{A deformation engineered so that the curvature-driven modulation of the
density of states can be measured separately from the pseudo Landau levels of
the lattice field} would decide how much of the observed response the
geometry is entitled to claim.

% ---------------------------------------------------------------------------
\ack
AR acknowledges support from Consejo de la Investigaci\'on Cient\'{\i}fica of
the Universidad Michoacana de San Nicol\'as de Hidalgo under grant 18371.
%\edit{Note for the authors: add support from Universidad del B\'{\i}o-B\'{\i}o
%and any acknowledgement DV considers appropriate.}

% ---------------------------------------------------------------------------
\appendix
\section{Gamma matrix conventions and the identities used in the text}
\label{app:gammas}

Signs are fixed once and for all by $\varepsilon^{01}=\varepsilon^{012}
=\varepsilon^{0123}=+1$ and by the mostly minus metric. In every
dimensionality the gamma matrices satisfy
$\{\gamma^{\hat\alpha},\gamma^{\hat\beta}\}=2\eta^{\hat\alpha\hat\beta}I$,
where the identity is \rev{$2\times2$} for $n=1$ and $n=2$ and
\rev{$4\times4$} for $n=3$.

On a line, with $\gamma^{0}=\sigma^{z}$ and $\gamma^{1}=i\sigma^{x}$, the
chirality matrix is \rev{$\gamma^{5}=\gamma^{0}\gamma^{1}=-\sigma^{y}$, which
satisfies $(\gamma^{5})^{2}=I$ and anticommutes with both gamma matrices.
The product $i\gamma^{0}\gamma^{1}$ squares to $-I$ and is not a chirality
matrix in this signature.} The single independent commutator is
\rev{$[\gamma^{0},\gamma^{1}]=2\gamma^{5}$}, equivalently
\rev{$4\Sigma^{\hat\alpha\hat\beta}=2\varepsilon^{\hat\alpha\hat\beta}
\gamma^{5}$}. Evaluating the components one by one confirms equation
(\ref{eq:dual1d}), \rev{and also that in $1+1$ dimensions the product
$\gamma^{\hat\alpha}\Sigma^{\hat\beta\hat\gamma}$ equals the metric term
alone}.

On a surface the third matrix is $\gamma^{2}=i\sigma^{y}$. The spinor keeps
\rev{two} components, so the anticommutator closes on $I_{2\times2}$, and the
commutators are those of equation (\ref{eq:clifford2d}). Equation
(\ref{eq:master2d}) then follows by writing
$\Sigma^{\hat\beta\hat\gamma}
=-\frac{i}{2}\varepsilon^{\hat\beta\hat\gamma\hat\lambda}
\gamma_{\hat\lambda}$ and using the Clifford algebra once.

In \rev{four} spacetime dimensions the required input is the product of
\rev{three} gamma matrices,
\begin{equation}
\gamma^{\hat\mu}\gamma^{\hat\nu}\gamma^{\hat\rho}
=\eta^{\hat\mu\hat\nu}\gamma^{\hat\rho}
+\eta^{\hat\nu\hat\rho}\gamma^{\hat\mu}
-\eta^{\hat\mu\hat\rho}\gamma^{\hat\nu}
-i\,\varepsilon^{\hat\sigma\hat\mu\hat\nu\hat\rho}
\gamma_{\hat\sigma}\gamma^{5}.
\label{eq:triple}
\end{equation}
Antisymmetrizing the last \rev{two} indices and dividing by \rev{four} turns
(\ref{eq:triple}) into the master identity (\ref{eq:master}), since the
metric terms combine in pairs while the Levi-Civita term doubles. Relation
(\ref{eq:triple}) is equivalent to the self duality of
$\sigma^{\hat\mu\hat\nu}
=\frac{i}{2}[\gamma^{\hat\mu},\gamma^{\hat\nu}]$ under multiplication by
$\gamma^{5}$.

\rev{Every identity in this appendix, the reductions (\ref{eq:Qsplit}) and
(\ref{eq:Q2d}), equations (\ref{eq:adiv}), (\ref{eq:symmetrized}),
(\ref{eq:rotation}), (\ref{eq:divK}) and the backgrounds quoted in
section~\ref{sec:applications} were checked symbolically in the matrix
representations quoted above, by building the spin connection from the frame
and comparing $Q$ computed directly with the gamma matrices against the
closed forms.} %\edit{Note for the authors: decide whether to deposit the
%verification and figure scripts in a public repository and cite it here.}

% ---------------------------------------------------------------------------
\section*{References}
%\bibliographystyle{iopart-num}
%\bibliography{refs}

\providecommand{\newblock}{}

\end{document}